\documentclass[twocolumn,showpacs,preprintnumbers,nofootinbib,amsmath,amssymb]{revtex4}

\usepackage{euscript,amssymb,amsmath}

\newcommand{\const}{\mbox{\rm const}\,}

\usepackage{graphicx}
\usepackage{epsfig}

\newcommand{\be}[1]{\begin{equation}\label{#1}}
\newcommand{\ee}{\end{equation}}
\newcommand{\ba}[1]{\begin{eqnarray}\label{#1}}
\newcommand{\ea}{\end{eqnarray}}
\newcommand{\rf}[1]{(\ref{#1})}
\newcommand{\nn}{\nonumber}
\newcommand{\epsch}{\bar\varepsilon_{\mathrm{Ch}}}
\newcommand{\epschnow}{\bar\varepsilon_{\mathrm{Ch,0}}}

\begin{document}

\title{Perfect fluids coupled to inhomogeneities in the late Universe}

\author{Alexander Zhuk}\email{ai.zhuk2@gmail.com}

\affiliation{ Astronomical Observatory,
 Odessa National University, Street Dvoryanskaya 2, Odessa 65082, Ukraine}


\begin{abstract}
We consider the Universe at the late stage of its evolution and deep inside the cell of uniformity.
At such scales, the Universe is highly inhomogeneous and is filled with inhomogeneities in the form of galaxies and the groups of galaxies. We also suggest that the Universe is filled with a perfect fluid, and its fluctuations have peculiar velocities of the same (non-relativistic) order of magnitude as for the inhomogeneities. In this sense, the inhomogeneities (e.g. galaxies) and fluctuations of perfect fluids are coupled with each other. We clarify some important points of this approach and present a brief review of previous studies (e.g. CPL model and Chaplygin gas). We demonstrate that considered perfect fluids which satisfy our approach are really coupled to galaxies concentrating around them. The averaged (over the whole Universe) value of their fluctuations is equal to zero.
\end{abstract}


\maketitle

\vspace{.5cm}


\section{\label{sec:1}Introduction}

\setcounter{equation}{0}

The modern observations clearly indicate that our Universe is dark. Nearly 96\%
of the mass-energy balance consists of dark energy and dark matter. However, the nature of these two
constituents is still unclear. To explain them, it was proposed a rather big number of theories where these dark sides were modeled by
different perfect fluids. The cosmological constant with the equation of state parameter $w=-1$ is the simplest example of it. The accelerating expansion of the late Universe is usually studied on the homogeneous and isotropic Friedmannian background.
On the other hand, deep inside the scale of homogeneity (of the order of 150-370 Mpc \cite{Yadav,Labini,EZcosm1})
the Universe is filled with discrete inhomogeneities in the form of galaxies, groups and clusters of galaxies. These inhomogeneities disturb the background Friedmann Universe which can be filled also with mentioned above perfect fluids.

The observations show that in the late Universe the peculiar velocities of galaxies and the groups and clusters of galaxies are highly non-relativistic. It is of interest to investigate a possibility that fluctuations of perfect fluids also acquire non-relativistic peculiar velocities, i.e. they are of the order of the peculiar velocities of dust-like matter (e.g. galaxies). In this sense, the inhomogeneities in the form of galaxies and fluctuations of perfect fluids are coupled with each other. Really, we shall show that if this happens, the perturbations of the energy density of perfect fluids can be concentrated around the galaxies screening their gravitational potentials.
This is a very specific form of perfect fluids. However, they can play an important role (e.g. in the form of dark matter) distributing around galaxies in such a way that to solve the problem of flatness of the rotation curves \cite{Laslo1,Laslo2}.

The main aim of this article is to clarify the notion of coupled fluids. To do it, we review some previous results and demonstrate which of perfect fluids can satisfy our approach and which not. We show that satisfactory perfect fluids are really coupled to galaxies concentrating around them.  In Sec. II, we briefly describe the theory of scalar perturbations for the coupled perfect fluids. In Sec. III, we consider perfect fluids with the constant negative equation of state (EoS) parameter $w=\const<0$. Then, in Sec. IV, we demonstrate that the CPL parametrization of the EoS cannot represent the coupled perfect fluid.
Sec. V is devoted to the Chaplygin gas model. Then, in Sec. VI, the total gravitational potential of the system is obtained for a special type of the coupled fluids. The main results are summarized in the concluding Sec. VII.




\section{\label{sec:2}Scalar perturbations}

\setcounter{equation}{0}

We consider the Universe at late stage of its evolution. We suppose that Universe is filled with pressureless matter (baryonic and dark), radiation, the cosmological constant and a perfect fluid. The latter can represent dark energy. This is not necessarily the case. For this reason, we have included also the cosmological constant. At distances being much larger than the scales of homogeneity, the Universe is well described by the homogeneous and isotropic FLRW metric
\be{2.1}
ds^2 = a^2(d\eta^2-\gamma_{\alpha\beta}dx^{\alpha}dx^{\beta})
\ee
and the following Friedmann equation
\ba{2.2}
&{}&\frac{2}{a^2}\left(\mathcal{H}'-\mathcal{H}^2-\mathcal{K}\right)= \nn\\
&-&\kappa\left(\overline{T}^0_{0\mathrm{\, (CDM)}}+\overline
\varepsilon_{\mathrm{rad}} +\overline \varepsilon_{\mathrm{X}} + \overline p_{\mathrm{rad}}+ \overline p_{\mathrm{X}}\right)\, ,
\ea
where ${\mathcal H}\equiv a'/a\equiv (da/d\eta)/a$, \ $\kappa\equiv 8\pi G_N/c^4$ ($c$ is the speed of light and $G_N$ is the Newtonian gravitational constant)
and $\mathcal K=-1,0,+1$ stands for open, flat and closed Universes, respectively. Conformal time $\eta$ and synchronous time $t$ are connected as $cdt=a d\eta$.
Further, $\overline T^{i}_{k\mathrm{\, (CDM)}}$ is the average energy-momentum tensor of the pressureless (dust-like) matter. This matter consists both visible and dark components in spite of the notation "CDM". For such matter the energy density
$\overline T^{0}_{0\mathrm{\, (CDM)}} =\overline \rho_{\mathrm{\, c}} c^2/a^3$ is the only nonzero component, here $\overline \rho_{\mathrm{\, c}}=\mbox{const}$
is the average comoving rest mass density \cite{EZcosm1}. As usual, for radiation we have the EoS $\overline p_{\mathrm{rad}}=(1/3)\overline
\varepsilon_{\mathrm{rad}}$. A barotropic perfect fluid has the equation of state (EoS) $\overline p_{\mathrm{X}}=\overline
p_{\mathrm{X}}(\overline \varepsilon_{\mathrm{X}})$.

Deep inside the cell of uniformity the Universe is highly inhomogeneous. Here, to describe the dynamical behavior of inhomogeneities, we can apply mechanical approach proposed in
\cite{EZcosm1,EZcosm2,EKZ2} (see also \cite{Novak} for the generalization to nonlinear $f(R)$ gravity). In the framework of the mechanical approach galaxies, dwarf galaxies and clusters of galaxies (composed of baryonic and dark
matter) are considered as separate compact objects. Moreover, at distances much greater than their characteristic sizes they can be described well as
point-like matter sources. This is a generalization of the well-known astrophysical approach \cite{Landau} (see \S 106) for the case of the dynamical
cosmological background. Usually, the gravitational fields of galaxies are weak and their peculiar velocities are much smaller than the speed of light.
These inhomogeneities together with fluctuations of other matter sources result in scalar perturbations of the FLRW metric:
\be{2.3}
ds^2 = a^2\left[(1+2\Phi)d\eta^2-(1-2\Phi)\gamma_{\alpha\beta}dx^{\alpha}dx^{\beta}\right]\, .
\ee
In the mechanical approach, the gravitational potential satisfies the following system of equations \cite{EZcosm2}:
\ba{2.4} &{}&\Delta\Phi-3\mathcal{H}(\Phi'+\mathcal{H}\Phi)+3\mathcal{K}\Phi\nn \\
&=&\frac{1}{2}\kappa a^2\left(\delta T^0_{0\mathrm{\, (CDM)}}+
\delta\varepsilon_{\mathrm{X}}+
\delta\varepsilon_{\mathrm{rad}}\right)\, ,\\
\label{2.5}
&{}&\frac{\partial}{\partial x^\beta}(\Phi'+\mathcal{H}\Phi)=0\, ,\\
\label{2.6} &{}&\Phi''+3\mathcal{H}\Phi'+(2\mathcal{H}'+\mathcal{H}^2)\Phi-\mathcal{K}\Phi\nn \\
&=&\frac{1}{2}\kappa a^2\left(\delta p_{\mathrm{X}}+ \delta
p_{\mathrm{rad}}\right)\, , \ea
where the Laplace operator $\triangle$ is defined with respect to the conformal spatial metric. Here, we took into account that all types of fluids are coupled to the inhomogeneities (i.e. galaxies). This means that the peculiar velocities of their fluctuations are the same order as peculiar velocities of the inhomogeneities. For this reason, the right hand side (r.h.s.) of Eq. \rf{2.5} is equal to zero (see details in \cite{EZcosm2}). This is the crucial point. Then, from Eq. \rf{2.5} we get immediately that
\be{2.7}
\Phi(\eta,{\bf r})=\frac{\varphi({\bf r})}{c^2a(\eta)}\, ,
\ee
where $\varphi({\bf r})$ is a function of all comoving spatial coordinates, and we have introduced $c^2$ in the denominator for convenience. In the surrounding of
an inhomogeneity, the comoving potential has an asymptotic behavior (see e.g. Sec. VI): $\varphi({\bf r})\sim 1/r$ for $r\to 0$, and the nonrelativistic gravitational potential $\Phi(\eta,{\bf
r})\sim 1/(a r)=1/R$, where $R=ar$ is the physical distance. Hence, $\Phi$ has the correct Newtonian limit near the inhomogeneities.
As we shall see additionally, in the most physically interesting cases fluctuations of the energy density of the coupled perfect fluids $\delta\varepsilon_{\mathrm{X}}$ are concentrated around inhomogeneities. Therefore, these perfect fluids are really coupled.

Obviously (see e.g. \cite{CPL}),
\be{2.8}
\delta p_{\mathrm{rad}} = \frac13 \delta \varepsilon_{\mathrm{rad}} \quad \Rightarrow \quad \delta \varepsilon_{\mathrm{rad}}\sim \frac{1}{a^4}
\ee
and from the definition of the EoS of a barotropic perfect fluid we also get
\be{2.9}
\delta p_{\mathrm{X}}=\frac{\partial \overline p_{\mathrm{X}}}{\partial \overline \varepsilon_{\mathrm{X}}}\delta \varepsilon_{\mathrm{X}}\, .
\ee

\vspace*{0.5mm}

Concerning the fluctuations of the matter sources, $\delta T^0_{0\mathrm{\, (CDM)}}$, it can be easily seen from Eq. (2.23) in \cite{EZcosm1} that these fluctuations read $\delta T^0_{0\mathrm{\,
(CDM)}}=\delta\rho_{\mathrm{c}}\, c^2/a^3 + (\overline{\rho}_{\mathrm{c}}c^2/a^3)[3\Phi +\alpha\Phi^2+\beta\Phi^3+...] $ where $\alpha$ and $\beta$ are some
constants. Taking into account Eq. \rf{2.7}, we get $\Phi/a^3 \sim O(1/a^4)$, $\Phi^2/a^3 \sim O(1/a^5)$, $\Phi^3/a^3 \sim O(1/a^6)$ and so on. Since we consider
linear perturbations, we should drop the terms $\Phi^2, \Phi^3, \ldots\; $. In what follows, we will keep terms up to the order $O(1/a^4)$ inclusive. Therefore,
\be{2.10}
\delta T^0_{0\mathrm{\, (CDM)}}=\frac{\delta\rho_{\mathrm{c}}\, c^2}{a^3}+\frac{3\overline{\rho}_{\mathrm{c}}\, c^2\Phi}{a^3}\, ,
\ee
where $\delta\rho_{\mathrm{\, c}}$ is the difference between the real and average comoving rest mass densities: $\delta\rho_{\mathrm{ c}} = \rho_{\mathrm{\,
c}}-\overline\rho_{\mathrm{\, c}}$.

Let us now analyze Eqs. \rf{2.4} and \rf{2.6}. Taking into account Eqs.~\rf{2.7}, \rf{2.2} and \rf{2.10}, we can rewrite them as follows:
\ba{2.11} &{}&\Delta\Phi+3\mathcal{K}\Phi\nn \\
&=& \frac{1}{2}\kappa a^2\left(\frac{\delta\rho_{\mathrm{c}}\,
c^2}{a^3}+\frac{3\overline\rho_\mathrm{c}c^2\Phi}{a^3}+\delta\varepsilon_{\mathrm{X}}+\delta\varepsilon_{\mathrm{rad}}\right)\,
\ea
and
\be{2.12}
-\left(\frac{\overline \rho_{\mathrm{ c}}\, c^2}{a^3}+\overline \varepsilon_{\mathrm{rad}} +\overline
\varepsilon_{\mathrm{X}} + \overline p_{\mathrm{rad}}+ \overline p_{\mathrm{X}}\right)\Phi
= \delta p_{\mathrm{X}}+\delta p_{\mathrm{rad}}\, .
\ee
Since according to the accuracy of our investigation we should keep terms up to $O(1/a^4)$ inclusive, the terms $\overline \varepsilon_{\mathrm{rad}} \Phi$ and $\overline p_{\mathrm{rad}} \Phi$ should be dropped from the latter equations because they are of the order $O(1/a^5)$:
\be{2.13}
-\left(\overline
\varepsilon_{\mathrm{X}} + \overline p_{\mathrm{X}}\right)\Phi
= \delta p_{\mathrm{X}}+ \frac{\overline \rho_{\mathrm{ c}} c^2 \Phi}{a^3} + \delta p_{\mathrm{rad}}\, .
\ee

In what follows, it is convenient to split $\delta\varepsilon_{\mathrm{rad}}$ into two parts:
\be{2.14}
\delta\varepsilon_{\mathrm{rad}} \equiv \delta\varepsilon_{\mathrm{rad1}}+\delta\varepsilon_{\mathrm{rad2}}\, ,
\ee
where
\be{2.15}
\delta\varepsilon_{\mathrm{rad1}}\equiv -\frac{3\overline \rho_{\mathrm{ c}} \varphi}{a^4}\, .
\ee
Then, Eqs. \rf{2.11} and \rf{2.13} read:
\be{2.16}
\triangle\varphi+3\mathcal{K}\varphi-\frac{\kappa c^4}{2}\delta\rho_{\mathrm{ c}}
=\frac{\kappa c^2 a^3}{2}\delta\varepsilon_{\mathrm{X}}+\frac{\kappa
c^2a^3}{2}\delta\varepsilon_{\mathrm{rad2}}\,
\ee
and
\be{2.17}
-\left(\overline{\varepsilon}_{\mathrm{X}}+\overline{p}_{\mathrm{X}}\right)\frac{\varphi}{c^2a}=
\frac{\partial \overline p_{\mathrm{X}}}{\partial \overline \varepsilon_{\mathrm{X}}}\delta \varepsilon_{\mathrm{X}}
+\frac{1}{3}\delta\varepsilon_{\mathrm{rad2}}\, , \ee
where we used Eqs. \rf{2.7}-\rf{2.9}. These are our master equations. Obviously,  the case $\overline p_{\mathrm{X}}=0$ is excluded here because it corresponds to the dust-like matter which was already taken into account in these equations. Already here, we arrive at one important conclusion.
The l.h.s. of Eq. \rf{2.16}  does not depend on time. Therefore, the same should hold true for its r.h.s. Hence, a perfect fluid can be coupled if it satisfies this condition.

Let us now investigate some of the popular models of perfect fluids to check whether they can be coupled fluids or not.


\section{\label{sec:3}Perfect fluid with the constant negative EoS parameter}

First, we consider a perfect fluid with the constant negative parameter of EoS:
\be{2.18}
\overline p_{\mathrm{X}} = {w} \overline \varepsilon_{\mathrm{X}}\, , \quad {w} =\const <0\, .
\ee
From the conservation equation we have
\be{2.19}
\overline \varepsilon_{\mathrm{X}} = \left(\frac{a_0}{a}\right)^{3(1+{w})}\overline \varepsilon_{\mathrm{X,0}}\, ,
\ee
where $a_0$ and $\overline\varepsilon_{\mathrm{X,0}}$ are the scale factor and the energy density of the considered perfect fluid at the present time.
Further, for fluctuations of this matter we have
\be{2.20}
\delta p_{\mathrm{X}} = {w} \delta\varepsilon_{\mathrm{X}}\, .
\ee
Therefore, Eq. \rf{2.17} gives
\be{2.21}
\delta\varepsilon_{\mathrm{X}} = -\frac{\varphi}{c^2a}\frac{1+{w}}{{w}}\bar\varepsilon_{\mathrm{X}} - \frac{1}{3{w}}
\delta\varepsilon_{\mathrm{rad2}}\, .
\ee
Substituting this relation into \rf{2.16}, we arrive at the following equation:
\ba{2.22}
&{}&\triangle\varphi+3\mathcal{K}\varphi-\frac{\kappa c^4}{2}\delta\rho_{\mathrm{ c}}\nn \\
&=&
-\frac{\kappa \varphi}{2}\frac{1+{w}}{{w}}a^2\bar\varepsilon_{\mathrm{X}}
+\frac{\kappa
c^2a^3}{2}\left(1-\frac{1}{3{w}}\right)\delta\varepsilon_{\mathrm{rad2}}\, .
\ea
We should determine now conditions which provide the time independence of the r.h.s. of this equation.
Because $\delta\varepsilon_{\mathrm{rad2}} \sim 1/a^4$, we must demand that
$\delta\varepsilon_{\mathrm{rad2}}\equiv 0$ (there is no possibility that the term with $\bar\varepsilon_{\mathrm{X}}$ and the term with
$\delta\varepsilon_{\mathrm{rad2}}$ cancel each other). Therefore, there are only two values of ${w}$ which do not contradict this equation. They are ${w} =-1$
and ${w} = -1/3$. The former case reduces to the standard $\Lambda$CDM model considered in \cite{EZcosm1}. The latter case ${w}=-1/3$ was considered in detail in
\cite{BUZ1} in the absence of radiation and briefly in \cite{CPL} in the presence of radiation. Therefore, if we exclude the $\Lambda$CDM case{\footnote{For the $\Lambda$CDM model (i.e. $w=-1$) fluctuations are absent, as it is easily seen from Eq. \rf{2.21}
where $\delta\varepsilon_{\mathrm{rad2}}=0$, so dark energy represented by the cosmological constant $\Lambda$ is truly homogeneous. Here, we cannot speak about coupling between the inhomogeneities in the form of galaxies and fluctuations of dark energy. Therefore, in spite of the fact that the $\Lambda$CDM model satisfies our approach, we cannot call it the coupled fluid in the accepted above sense.}}, there is only one possibility for the considered fluid to be coupled. This is the $w=-1/3$ case.
Then, the fluctuation of the perfect fluid reads
\be{2.22a}
\delta\varepsilon_{\mathrm{X}}=
2{\left(\frac{a_0}{a}\right)^{2}
\frac{\overline\varepsilon_{\mathrm{X,0}}}{c^2 a}}\varphi
\, .
\ee


\section{\label{sec:4}CPL and other models with linear parametrization of EoS}

In the CPL model (i.e. $\mathrm{X} \equiv \mathrm{CPL}$) \cite{ChevPol,Linder} the parameter of EoS is the linear function with respect to the scale factor of the Universe:
\ba{2.23}
\overline p_{\mathrm{CPL}}={w}(a)\overline\varepsilon_{\mathrm{CPL}},\, \,
{w}(a)&=&{w}_{0}+{w}_{1}\left(1-\frac{a}{a_{0}}\right),\\
 {w}_1 \neq 0\, ,&{}&\nn
\ea
where $a_0$ represents the scale factor of the Universe at the present time. We consider this EoS to be valid for a definite period of time in the past
(e.g., from last scattering \cite{Linder}). We also suppose that this EoS is still valid for some period of time in the future.

From the conservation equation  we easily get
\be{2.24}
\overline\varepsilon_{\mathrm{CPL}}=Aa^{-3\left(1+{w}_{0}+{w}_{1}\right)} e^{3{w}_{1}a/a_{0}} \, ,
\ee
where $A$ is a constant of integration. With the help of this equation and definition \rf{2.9}, we obtain the expression for the fluctuation of the CPL fluid pressure:
\ba{2.25}
&{}&\delta p_{\mathrm{CPL}}
=\left({w}-\frac{{w}_1\overline\varepsilon_{\mathrm{CPL}}}{a_0}\frac{da}{d\overline\varepsilon_{\mathrm{CPL}}}\right)
\delta\varepsilon_{\mathrm{CPL}} \\
&=& \left[{w}_{0}+{w}_{1}\left(1-\frac{a}{a_{0}}\right)-
\frac{1}{3}\frac{{w}_1}{{w}_1-\frac{a_0}{a}\left(1+{w}_{0}+{w}_{1}\right)}\right]\delta\varepsilon_{\mathrm{CPL}}\, .\nn
\ea
Then, we can determine $\delta\varepsilon_{\mathrm{CPL}}$ from Eq. \rf{2.17}:
\ba{2.26}
&{}&\delta\varepsilon_{\mathrm{CPL}} =
-\frac{A\varphi}{c^2}a^{-3\left(1+{w}_{0}+{w}_{1}\right)-1}
e^{3{w}_{1}\beta} \nn \\
&\times&\frac{\left(1+{w}_{0}+{w}_1\left(1-\beta\right)\right)^2}{{w}_{0}+{w}_1\left(1-2\beta/3\right)+\left({w}_{0}+
{w}_1\left(1-\beta\right)\right)^2} \\
&-& \frac{1}{3}\delta\varepsilon_{\mathrm{rad2}}\frac{1+{w}_{0}+{w}_1\left(1-\beta\right)}{{w}_{0}+{w}_1\left(1-2\beta/3\right)+\left({w}_{0}+
{w}_1\left(1-\beta\right)\right)^2}\nn \, ,
\ea
where for convenience we have introduced the notation $\beta \equiv a/a_0$. Consequently, Eq.~\rf{2.16} takes the form
\ba{2.27}
&{}&\triangle\varphi+3\mathcal{K}\varphi-\frac{\kappa c^4}{2}\delta\rho_{\mathrm{ c}} =-\frac{A \kappa \varphi}{2}a^{-3\left(1+{w}_{0}+{w}_{1}\right)+2}
e^{3{w}_{1}\beta}\nn \\
&\times&\frac{\left(1+{w}_{0}+{w}_1\left(1-\beta\right)\right)^2}{{w}_{0}+{w}_1\left(1-2\beta/3\right)+\left({w}_{0}+
{w}_1\left(1-\beta\right)\right)^2}
+\frac{\kappa c^2a^3}{2}\delta\varepsilon_{\mathrm{rad2}}\nn\\
&\times&\left[
1-\frac{1}{3}\frac{1+{w}_{0}+{w}_1\left(1-\beta\right)}{{w}_{0}+{w}_1\left(1-2\beta/3\right)+\left({w}_{0}+
{w}_1\left(1-\beta\right)\right)^2}\right]\, .
\ea

Since the l.h.s. of this equation does not depend on time, either each of two
remaining expressions are also independent of time (within our accuracy $O(1/a)$, bearing in mind that the r.h.s. of \rf{2.27} has been multiplied by $a^3$), or they
must cancel each other at any arbitrary moment of time. Can this be the case? The second assumption that the second and third terms on the r.h.s. of Eq.~\rf{2.27} can
cancel each other at any arbitrary moment of time does not work because of the presence of the exponential (with respect to the scale factor $a$) function
$\exp({3{w}_{1}\beta})$ which is linearly independent from the power functions. It can be easily seen that the first assumption does not work as well in the case ${w}_1\neq 0$ (see the details in \cite{CPL}). Therefore, perfect fluids with the CPL parametrization cannot be coupled fluids.

By performing an analysis similar to the one carried above in this section, we conclude that models with the following linear parametrization of EoS:
\ba{2.28}
&{}&w(z)=w_{0}-w_{1}z=w_{0}+w_{1}\left(1-\frac{a_{0}}{a}\right)\, ,\\
&{}&w(t)=w_{0}+w_{1}\left(1-\frac{t}{t_{0}}\right)\label{2.29}
\ea
also cannot represent the coupled fluids \cite{CPL}.


\section{\label{sec:5}Chaplygin gas model}

The barotropic perfect fluid called the modified generalized Chaplygin gas (mGCG) (here, $X\equiv$Ch) is characterized by the equation of state (EoS) \cite{Benaoum:2002zs}
\be{2.30}
	\overline p_{\mathrm{Ch}}
		= \beta \epsch
		- (1+\beta)\frac{A}{\bar\varepsilon^{\alpha}_{\mathrm{Ch}}}\, ,\quad \beta,\alpha \neq -1\,  \mathrm{and}\,  A\neq 0\, ,
\ee
where $\beta$, $A$ and $\alpha$ are the constant parameters of the model.
Then, the conservation equation results in the following dependence of the energy density on the scale factor $a$:
\be{2.31}
	\epsch
		=\epschnow\left[
			A_s
			+\left(1-A_s\right)\left(\frac{a_0}{a}\right)^{3\xi}
		\right]^{\frac{1}{1+\alpha}}\, ,
\ee
where
\be{2.32}
\xi \equiv (1+\beta)(1+\alpha)
\ee
and $a_0$ and $\epschnow$ are the current values of the scale factor and the energy density of the mGCG, and $A_s\equiv A/\epschnow^{1+\alpha}$. Here, we exclude the case $A_s=0$, as it corresponds to $A=0$, and the case $A_s=1$ because it reduces to the trivial cosmological constant case. A simple analysis shows that the late time acceleration of the Universe
takes place if $\xi >0$, and the Chaplygin gas behaves asymptotically as the cosmological constant. In what follows, we consider only this case.

According to the formula \rf{2.9}, we get from Eq. \rf{2.31}
\be{2.33}
	\delta p_{\mathrm{Ch}}
		=\frac{
			A_s(\xi-1) + (1-A_s)\beta (\frac{a_0}{a})^{3\xi}
		}
		{
			A_s + (1-A_s) (\frac{a_0}{a})^{3\xi}
		}
		\delta\varepsilon_{\mathrm{Ch}}
	\, .
\ee
Then, taking into account Eqs.~\rf{2.30} and \rf{2.31}, we get from Eq.~\rf{2.17} the expression for $\delta\varepsilon_{\mathrm{Ch}}$:
\ba{2.34}
	\delta\varepsilon_{\mathrm{Ch}}
		&=& -
		\left[
			A_s
			+ (1-A_s)\left(\frac{a_0}{a}\right)^{3\xi}
		\right]^{\frac{1}{1+\alpha}}\nn\\
&\times&
		\frac{(1-A_s)(1+\beta)(\frac{a_0}{a})^{3\xi}}{A_s(\xi-1) + (1-A_s)\beta (\frac{a_0}{a})^{3\xi}}
		\frac{\epschnow}{c^2  a}\varphi \\
		&{ }&-\frac{1}{3}
		\frac{
			A_s
			+ (1-A_s)\left(\frac{a_0}{a}\right)^{3\xi}
		}
		{
			A_s(\xi-1) + (1-A_s)\beta\left(\frac{a_0}{a}\right)^{3\xi}
		}
		\delta\varepsilon_{\mathrm{rad2}}
		\, .\nn
\ea

Now, we can investigate the consistency of Eq.~\rf{2.16} with $\delta\varepsilon_{\mathrm{Ch}}$ given in Eq.~\rf{2.34} for different values of $\xi$.
As we mentioned above, the system of Eqs.~\rf{2.16} and \rf{2.17} was derived for the late Universe, i.e. for rather big values of the scale factor $a$, starting e.g. from the present moment. To get these equations, we neglect terms $o(1/a^4)$. This means that fluctuation $\delta\varepsilon_{\mathrm{X}}$ should be defined up to $O(1/a^4)$ inclusive. Obviously, the larger the value of $a$, the more accurate our mechanical approach is.
That is, it should work perfectly in the future Universe. On the other hand, Eq.~\rf{2.34} and consequently \rf{2.16} are considerably simplified in the approximation of big values of the scale factor, $a$, as we are neglecting terms of the order $o(1/a^{4})$. Following this prescription, we first expand the expression \rf{2.34} in powers of $1/a$, neglecting terms $o(1/a^4)$. Then we substitute this expression into Eq.~\rf{2.16} and analyse the consistency of the equation obtained. The equation is considered consistent if its r.h.s. does not depend on time up to terms $O(1/a)$ inclusive (bearing in mind that the r.h.s. of this equation has been multiplied by $a^3$). It is clear that the result depends on the values of the parameters of the model. If we can define the sets of parameters which make this equation consistent, then we could claim that the corresponding Chaplygin gas can represent the coupled fluid.

As we can see from Eq.~\rf{2.34}, the powers of the ratio $(1/a)$ are mainly defined by the parameter $\xi$. Therefore, it makes sense to investigate this equation separately for different values of $\xi$. As it follows from \cite{Chaplygin}, the most interesting physical case corresponds to $0<\xi<1$.
In this case, for big values of the scale factor, $a$, the mGCG fluctuation \rf{2.34} is approximated as
\ba{2.35}
	&{}&\delta\varepsilon_{\mathrm{Ch}}
	\approx
	\frac{1+\beta}{1-\xi}\frac{1-A_s}{A_s}A_s^{\frac{1}{1+\alpha}}
	{ \left(\frac{a_0}{a}\right)^{3\xi}
	\frac{\epschnow}{c^2 a}}
	\varphi\nn \\
	&+& \frac{1}{3(1-\xi)}\delta\varepsilon_{\mathrm{rad2}}
	\, ,
\ea
and after substituting it in Eq.~\rf{2.16} we get
\ba{2.36}
	&{}&\triangle\varphi+3\mathcal{K}\varphi-\frac{\kappa c^4}{2}\delta\rho_{\mathrm{c}}\nn \\
&=&
	\frac{1+\beta}{1-\xi}\frac{1-A_s}{A_s}A_s^{\frac{1}{1+\alpha}}
	{ \left(\frac{a_0}{a}\right)^{3\xi}
	\frac{\kappa a^2\epschnow}{2}}\varphi
	\nn\\
	&{+}&\left(
		1
		+ \frac{1}{3(1-\xi)}
	\right)
	{\frac{\kappa c^2 a^3}{2}}
	\delta\varepsilon_{\mathrm{rad2}}
\, .
\ea
Now, we need to find conditions under which the r.h.s. of this equation is independent of time.
Because $-2<3\xi-2<1$, the first term on the r.h.s. in Eq.~\eqref{2.36} cannot be compensated by the second one, which behaves as $\delta\varepsilon_{\mathrm{rad2}} \sim 1/a^4$. The only way to solve this problem is to put $\delta\varepsilon_{\mathrm{rad2}}\equiv 0$ and to make the first term in Eq.~\eqref{2.36} independent on time. The latter condition takes place if
$3\xi-2=0\, \Rightarrow \, \xi =2/3$. Then, the mGCG fluctuation reads
\be{2.37}
	\delta\varepsilon_{\mathrm{Ch}}=
	3(1+\beta)\frac{1-A_s}{A_s}A_s^{\frac{1}{1+\alpha}}
	{\left(\frac{a_0}{a}\right)^{2}
	\frac{\epschnow}{c^2 a}}\varphi
	\, .
\ee

\section{\label{sec:6}Gravitational potential}

In this section we consider one of the most physically interesting cases $\delta\varepsilon_{\mathrm{rad2}}=0$. Both models of coupled fluids found in sections III and V belong to this case. Then, for the fluctuations $\delta\varepsilon_{\mathrm{X}}$ we get from \rf{2.17}:
\be{2.39}
\delta \varepsilon_{\mathrm{X}} = -\left[ \left.\left(\overline{\varepsilon}_{\mathrm{X}}+\overline{p}_{\mathrm{X}}\right)\right/
\frac{\partial \overline p_{\mathrm{X}}}{\partial \overline \varepsilon_{\mathrm{X}}}\right]\frac{\varphi}{c^2a}\equiv
E_{\mathrm{X}}(a)\frac{\varphi}{c^2a}\neq 0
\, .
\ee
Here, we introduced a function $E_{\mathrm{X}}(a)$ which depends on the form of the EoS of a perfect fluid. The case $\overline p_{\mathrm{X}}=0$ was excluded from the consideration, and, since we consider non-zero fluctuations $\delta \varepsilon_{\mathrm{X}}$,
we also exclude the case $\overline{p}_{\mathrm{X}}=-\overline{\varepsilon}_{\mathrm{X}}$. Therefore, Eq. \rf{2.16} reads
\be{2.40}
\triangle\varphi+3\mathcal{K}\varphi-\frac{\kappa c^4}{2}\delta\rho_{\mathrm{ c}}=\frac{\kappa c^2 a^3}{2}E_{\mathrm{X}}(a)\frac{\varphi}{c^2a}\, .
\ee
Obviously, this equation is self-consistent only if the function $E_{\mathrm{X}}(a)$ has the following asymptotic behavior for big values of the scale factor $a$:
\be{2.41}
E_{\mathrm{X}}(a) = \frac{\overline A_{\mathrm{X}}}{a^2} +o(1/a^3)\, .
\ee
This is the necessary and sufficient condition for the coupled fluids in the considered case. For example, from Eqs. \rf{2.22a} and \rf{2.37} for the coupled perfect fluids with the parameter of the EoS $w=-1/3$ and Chaplygin gas, we have respectively:
\ba{2.42}
\overline A_{\mathrm{X}}&=&2 (a_0)^2 \overline \varepsilon_{\mathrm{X,0}}\, ,\\
\label{2.43}\overline A_{\mathrm{Ch}}&=&
3(1+\beta)\frac{1-A_s}{A_s}A_s^{\frac{1}{1+\alpha}}
	\left({a_0}\right)^{2}
	\epschnow\, .
\ea
It is worth  noting that in the former case, terms $o(1/a^3)$ in formula \rf{2.41} are absent.

Then, Eq. \rf{2.40} takes the form
\be{2.44}
\triangle\varphi+\left(3\mathcal{K} - \frac{\kappa\overline A_{\mathrm{X}}}{2}\right)\varphi=
\frac{\kappa c^4}{2}\left(\rho_{\mathrm{ c}}-\overline\rho_{\mathrm{ c}}\right)\, .
\ee
With the help of the substitution
\be{2.45}
\varphi = \phi + \lambda^2 \, \, \frac{\kappa c^4}{2}\overline\rho_{\mathrm{ c}}\, ,
\ee
where
\be{2.46}
\lambda^2 \equiv \left[\frac{\kappa \overline A_{\mathrm{X}}}{2}-3\mathcal{K}\right]^{-1}\, ,
\ee
Eq. \rf{2.44} takes the Hemholtz form:
\be{2.47}
\triangle\phi-\frac{1}{\lambda^2}\phi = \frac{\kappa c^4}{2}\rho_{\mathrm{ c}}\, ,
\ee
where the comoving rest mass density of the inhomogeneities (e.g. galaxies of the masses $m_i$) is \cite{EZcosm1,BUZ1}:
\be{2.48}
\rho_{\mathrm{ c}} = \frac{1}{\sqrt{\gamma}}\sum_i m_i \delta(\mathbf{r}-\mathbf{r}_i)\, .
\ee
Obviously, we first can find a solution for a single mass $m_i$ of an inhomogeneity and then apply the superposition principle to get a total solution.
Solutions of Eq. \rf{2.47} depend on the type of the spatial topology of the Universe, i.e. on the sign of $\mathcal{K}$.  They were investigated in detail in \cite{BUZ1} for any sign of $\mathcal{K}$. It is shown there that the coupled fluids result in the screening of the gravitational potentials of inhomogeneities. For example, in the case of the flat space $\mathcal{K}=0$, the solution for a single mass $m_i$ and for a proper boundary conditions is:
\be{2.49}
\phi_i=-\frac{G_Nm_i}{r}\exp(-r/\lambda)\, , \quad  \; 0<r<+\infty\, .
\ee
The total gravitational potential takes the form
\be{2.50}
\varphi=-G_N\sum_i\frac{m_i}{|{\bf r}-{\bf r}_i|}\exp\left(-|{\bf r}-{\bf r}_i|/\lambda\right)+ \lambda^2 \, 4\pi G_N\overline\rho_{\mathrm{c}}\, .
\ee

Eq. \rf{2.39} shows that the perfect fluid fluctuations are proportional to the total potential:
\be{2.51}
\delta \varepsilon_{\mathrm{X}} \propto \varphi\, .
\ee
This relation results to two important conclusions for the considered coupled fluids. First, such perfect fluids are concentrated around the
inhomogeneities (e.g. galaxies) screening their gravitational potentials. Therefore, these perfect fluids are really coupled to the inhomogeneities. Second, the averaged over the whole Universe fluctuations are connected with the averaged value of the total gravitational
potential: $\overline{\delta \varepsilon_{\mathrm{X}}} \propto \overline{\varphi}$. It is not difficult to show that  the averaged value of the total potential \rf{2.45} (where $\phi$ is the solution of \rf{2.47}) is equal to zero \cite{BUZ1}. It takes place for any sign of $\mathcal{K}$. Hence, the averaged value of fluctuation is also equal to zero: $\overline{\delta \varepsilon_{\mathrm{X}}}=0$. This is a physically reasonable result.


\section{Conclusions}

In the present paper we have considered our Universe at the late stage of its evolution and deep inside the cell of uniformity.
At such scales the Universe is highly inhomogeneous: we can clearly see the inhomogeneities in the form of galaxies and the group of galaxies. We have also suggested that the Universe is filled with a perfect fluid which can be dark energy. It was supposed that fluctuations of these perfect fluids and inhomogeneities are coupled to each other. These means that they have peculiar velocities of the same (non-relativistic!) order of magnitude. A theory of scalar perturbations for such coupled fluids was developed in \cite{EZcosm1,EZcosm2,CPL}. In the present paper, we have clarified some points of this approach and have reviewed previous results \cite{BUZ1,CPL,Chaplygin}. We have demonstrated that considered perfect fluids which satisfy our approach are really coupled to galaxies concentrating around them.
The averaged (over the whole Universe) value of their fluctuations is equal to zero. It is worth noting that minimal scalar field can also be coupled to galaxies \cite{scalar field 1}. The same concerns dark matter and dark energy, interacting with each other in the non-gravitational way \cite{EKief}.

It is of interest to generalize the mechanical approach to the case of non-zero peculiar velocities and for all cosmological scales. For the $\Lambda$CDM model and a specific perfect fluid model, such generalization was performed in the recent papers \cite{Ein1,Ein2}. It make sense to investigate also perfect fluids considered in our present paper and to see the transition from the more general case to the coupled fluid case.

\section*{ACKNOWLEDGEMENTS}

I would like to thank Maxim Eingorn for the useful discussion during the preparation of this paper. I also thank Mariam Bouhmadi-L\'opez, \"{O}zg\"ur Akarsu, Jo\~ao~Morais, Alvina Burgazli, Maxim Brilenkov and Ruslan Brilenkov for the fruitful collaboration during previous works which formed the basis of this mini review.



\end{document}